\documentclass[%
 reprint,
 amsmath,amssymb,
 aps,
]{revtex4-2}

\usepackage{colortbl}  
\usepackage{xcolor}    

\colorlet{mixColor}{blue!30!red!30}
\colorlet{mixColor1}{red!50!yellow!30}
\colorlet{mixColor2}{red!30!yellow!30}
\colorlet{mixColor3}{green!30!yellow!30}
\colorlet{mixColor4}{green!20!blue!20}
\usepackage{graphicx}
\usepackage{subcaption}
\usepackage{dcolumn}
\usepackage{bm}
\usepackage{quantikz} 
\usepackage{hyperref}
\usepackage{amsmath,mathrsfs}
\usepackage{soul,ulem}

\usepackage{orcidlink} 

\DeclareMathOperator{\Tr}{Tr}

\begin{document}

\preprint{APS/123-QED}

\title{Entangling power, gate typicality and Measurement-induced Phase Transitions}

\author{Sourav Manna\orcidlink{0009-0005-5216-567X}}
\email{mannaphy@physics.iitm.ac.in}

\author{Vaibhav Madhok\orcidlink{https://orcid.org/0000-0003-0785-0094}}%

\author{Arul Lakshminarayan\orcidlink{https://orcid.org/0000-0002-5891-6017}}
\email{arul@physics.iitm.ac.in}
\affiliation{
	Department of Physics and Center for Quantum Information, Communication and Computing,
	Indian Institute of Technology Madras, Chennai, India 600036
}

\date{\today}

\begin{abstract}
When subject to a non-local unitary evolution, qubits in a quantum circuit become increasingly entangled. Conversely, measurements applied to individual qubits lead to their disentanglement from the collective system. The extent of entanglement reduction depends on the frequency of local projective measurements. A delicate balance emerges between unitary evolution, which enhances entanglement, and measurements which diminish it. In the thermodynamic limit, there is a phase transition from volume law entanglement to area law entanglement at a critical value of measurement frequency. This phenomenon, occurring in hybrid quantum circuits with both unitary gates and measurements, is termed as measurement-induced phase transition (MIPT). We study the behavior of MIPT in circuits comprising of two-qubit unitary gates parameterized by Cartan decomposition. We show that the entangling power and gate typicality of the two-qubit local unitaries employed in the circuit can be used to explain the behavior of global bipartite entanglement the circuit can sustain. When the two-qubit gate throughout the circuit is the identity and measurements are the sole driver of the entanglement behavior, we obtain analytical estimate for the entanglement entropy that shows remarkable agreement with numerical simulations. 
We also find that the entangling power and gate typicality enable the classification of the two-qubit unitaries by different universality classes of phase transitions that can occur in the hybrid circuit.  For all unitaries in a particular universality class, the transition from volume to area law of entanglement occurs 
with same exponent that characterizes the phase transition. 
\end{abstract}

\maketitle


\section{\label{sec:intro}Introduction}
The interplay between complexity, entropy and entanglement in many-body systems has been crucial to the foundations of quantum statistical physics \cite{popescu2006entanglement}. For example, how quantum systems, isolated from the external environment, thermalize are being actively studied \cite{Deutsch_1991,Cazalilla_2010, rigol2008thermalization}.
The connections between multipartite entanglement in a circuit to quantum advantage or quantum ``volume''  and to efficient classical simulation of such circuits are central to quantum information processing. In particular, 
they address the question how exactly one can think of entanglement as a resource in quantum computation \cite{jozsa2003role, RevModPhys.80.517}.
Recently, discovery of measurement-induced phase transitions which involves an interplay of entanglement generation and local measurements that destroy the entanglement has 
added to the richness of emerging paradigms studied at the intersection of quantum statistical mechanics and quantum information science\cite{Skinner_2019, PhysRevB.101.060301, PhysRevResearch.4.043212, PhysRevX.11.011030}.

Consider a quantum circuit that has multi-qubit unitaries acting on its constituents. 
A circuit starting in a product state becomes entangled after such a unitary evolution. After a sufficiently long time the quantum state gains multipartite entanglement. For such a generic quantum circuit with non-local gates, the entanglement entropy scales as the system size, which is the number of qubits $L$, called volume law of entanglement \cite{CalabreseCardy2005, NahumPrX2017}. 
If entanglement entropy does not scale with system size, we call it as an area law of entanglement. Between these two phases of entanglement there is a critical value of entanglement where the entanglement scales as $\sim \ln{L}$ \cite{Calabrese_2004}. When we measure a qubit, it becomes unentangled from  rest of the qubits.
While it will therefore not contribute to the entanglement immediately after the measurement, subsequent dynamics will entangle it with other qubits and therefore we retain the measured qubits in the circuit.
Therefore, in a circuit, if one makes many local measurements on individual qubits with  high frequency, one collapses the collective wave function and does not let the global entanglement grow.

We consider a hybrid circuit with two operations, unitary evolution via circuit gates and quantum measurements that are interspersed \st{in} between them. Depending on the frequency of measurement, or equivalently the probability $p$ of a local projective measurement at each qubit, 
one can observe a phase transition from volume law to area law entanglement. This is referred to as measurement-induced phase transition (MIPT) \cite{Skinner_2019, Fisher_2023, Iadecola_2023, Li_2019, PhysRevB.102.224311, PhysRevLett.132.110403, NahumRoyPRX2021, BuchholdAltladPrx2021, LucaMIPT2024}. If $p_c$ is the critical probability, for $p< p_c$ the circuit will be in the volume law phase and for $p> p_c$ it will exhibit an area law phase. Earlier works show that  measurement induced phase transition occurs for Ising spin systems \cite{PhysRevB.102.094204, PhysRevB.102.035119, PhysRevB.108.165120, PhysRevB.103.224210, 10.21468/SciPostPhys.15.3.096, PhysRevB.108.184302}, Bose-Hubbard models\cite{PhysRevResearch.2.013022, PhysRevB.102.054302} and SYK models \cite{PhysRevLett.127.140601}.
Experimentally  such phase transition have been observed in superconducting quantum processors \cite{2023, Koh_2023} and ion-trap quantum computers \cite{Noel_2022}.

Phase diagrams give us information about the different states of a substance under different conditions of temperature and pressure. 
The motivation is to take the first steps in giving a phase diagram for measurement-induced phase transitions in an analogy with statistical physics.  
In particular, we address the following questions - how does the behavior of a measurement-induced phase transition depend on the properties of two-qubit unitaries employed in the quantum circuit? In particular, how does the value of the critical probability $p_c$ vary with the entangling power, or the interaction content of the two-qubit unitary building block  gates?  



For a given two qudit (not necesary restricted to qubits) unitary $U$ there is a well-defined measure, the entangling power $e_p(U)$,  \cite{Zanardi_2000, Zanardi_2001} that quantifies the average amount of entanglement generated by its action on product states.
It is straightforward to see that this quantity is a local unitary (LU) invariant, namely 
\begin{equation}
    e_p[(u_1\otimes u_2)U (v_1\otimes v_2)]=e_p(U)
\end{equation} where $u_i, v_i$ are any single qudit unitary operators. 
Here we consider brick-wall circuits constructed of different two-qubit unitaries, but constrained to  having the same entangling power. Thus they are of the form $U=(u_1\otimes u_2)\, \mathcal{U}\, (v_1\otimes v_2)$, where we fix the interaction part $\mathcal{U}$ for all two-qubit gates and allow Haar random sampled single qubit unitaries across both space and time. For a given probability of measurement we find the average entanglement entropy after a sufficiently long time and average this over the random single qubit unitaries. By construction this is an LU invariant quantity. The critical probability where the circuit  exhibits a transition from volume to area law of entanglement is then found from this average and is also LU invariant. Systematically sampling the interaction two-qubit unitary $\mathcal{U}$ over all possibilities, the critical probability $p_c$ is found as a function of known two-qubit LU invariants. 

 
 Heuristically, if the two-qubit building block gates have higher entangling power, we may expect a  faster  increase in entanglement and a higher critical probability of measurements to render it an area law. However, interestingly we find that two-qubit gates with the same entangling power result in different behavior for the state of the many-body quantum circuit's entanglement entropy. The critical parameters of the phase transition also change, indicating that other two-qubit gate properties are at work.
In this work, we identify entangling power along with the ``gate typicality'' \cite{Jonnadula_2017,Jonnadula_2020} as the two-qubit gate parameters that determine the behavior of entanglement entropy of the entire many-body circuit and its critical properties. This seems to 
classify the set of two-qubit unitaries into different universality classes of MIPT that they give rise to.

The strategy of having only the single qudit operators as random elements in circuits has been used previously, at least to study the growth of entanglement \cite{KuoArovasPrb2020, BensaPRX2021}, and in the study of correlation functions \cite{AravindaSuhail2021}. These ``locally scrambled" circuits are low in quantum resources, but also allow to study the optimal set of entangling gates for a given task. If the task is to generate entanglement, from the purity decay studies of \cite{BensaPRX2021}, the iSWAP or DCNOT gate is singled out as the one leading to the fastest rate. In this work, we find that with the inclusion of measurements, this continues to remain the case. We find the largest critical probabilities in these cases.

\section{\label{sec:model}Background: The circuit model, entangling power and gate typicality}

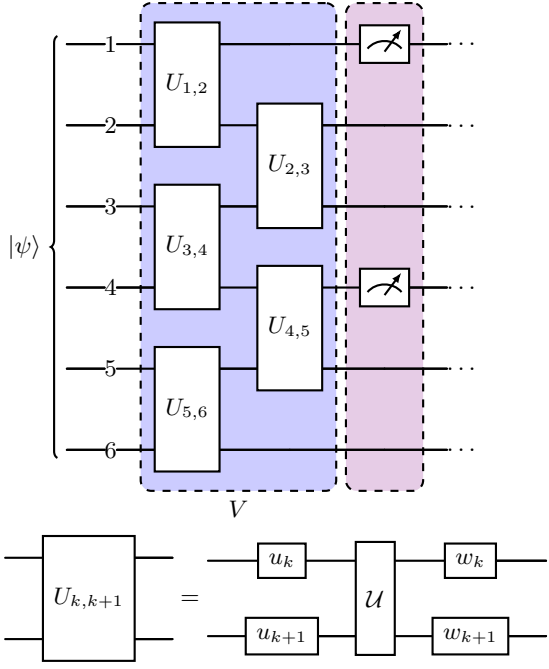
\begin{figure}[!htbp]
	\begin{subfigure}{0.5\textwidth}  
		\centering
		\begin{quantikz}
			\lstick[6]{\ket{\psi}} & \qw 1 & \gate[2]{U_{1,2}} \gategroup[6,steps=2,style={dashed,rounded corners,fill=blue!20, inner xsep=2pt},background,label style={label
				position=below,anchor=north,yshift=-0.2cm}]{$V$}& \qw & \meter{} \gategroup[6,steps=1,style={dashed,rounded corners,fill=violet!20, inner xsep=2pt},background]{} & \qw \ldots\\
			& \qw 2 & \qw &\gate[2]{U_{2,3}} & \qw & \qw \ldots\\
			& \qw 3 & \gate[2]{U_{3,4}} & \qw & \qw & \qw \ldots\\
			& \qw 4 & \qw & \gate[2]{U_{4,5}} &\meter{} & \qw \ldots\\
			& \qw 5 & \gate[2]{U_{5,6}} & \qw & \qw & \qw \ldots\\
			& \qw 6 & \qw & \qw & \qw & \qw \ldots
		\end{quantikz}
	\end{subfigure}\\

	\begin{subfigure}{0.55\textwidth}
		\centering
		\begin{quantikz}
			\qw &  \gate[2]{U_{k,k+1}} & \qw\\
			& & \qw
		\end{quantikz}=\begin{quantikz}
			& \gate[1]{u_k} & \gate[2]{\mathcal{U}} & \gate[1]{w_k} & \qw\\
			& \gate[1]{u_{k+1}} & \qw & \gate[1]{w_{k+1}} & \qw
		\end{quantikz}
	\end{subfigure}
	\caption[]{ The hybrid circuit with unitary gates interspersed with local measurements. The blue region is the unitary evolution, $V$, given in Eq.~\eqref{eq1.0} to generate entanglement. After unitary evolution, qubits are randomly measured with a probability in the violet layer. The structure of $U_{k,k+1}$, shown below the hybrid circuit, is $U_{k,k+1} = (w_k\otimes w_{k+1})\mathcal{U}(u_k\otimes u_{k+1})$. }
	\label{fig1}
\end{figure}

 Measurement-induced phase transitions have been studied for different quantum circuit models like dynamical quantum trees \cite{PRXQuantum.4.030333} and brick-wall  quantum circuits \cite{Skinner_2019, PhysRevB.101.104301, PhysRevX.10.041020}. In this study, we employ the brick-wall open chain circuit with random measurement as shown in Fig.~\ref{fig1}. Here $U$ is a two-qubit unitary matrix, the nearest-neighbor interaction. One step unitary evolution matrix, as shown in Fig.~\ref{fig1}, is 
\begin{equation}
    \label{eq1.0}
    V = \left(\prod_{j: \text{even}} U_{j,j+1}\right) \left(\prod_{i:\text{odd}} U_{i,i+1}\right).
\end{equation}

Between every unitary evolution step, $V$, which themselves are not identical, a random set of qubits are subjected to Pauli-$z$ measurements. At each time step a qubit is measured if an independent uniform random number generator in $[0,1]$ returns a value $\leq p$. Thus an average number $pL$ of qubits are measured between any two consecutive unitary evolutions.
Such circuits with unitary gates interspersed with local measurements have been called ``hybrid circuits'' \cite{Li_2019}, and we refer to them as such below. 
 We calculate the entanglement between the first $L/2$ qubits with the rest (we take $L$ even). Throughout this work, we use von Neumann entropy, $\mathcal{S} = -\Tr(\rho_A \ln{\rho_A})$, as the measure of bipartite entanglement between the two partitions. Here, $\rho_A$ is the subsystem density matrix.

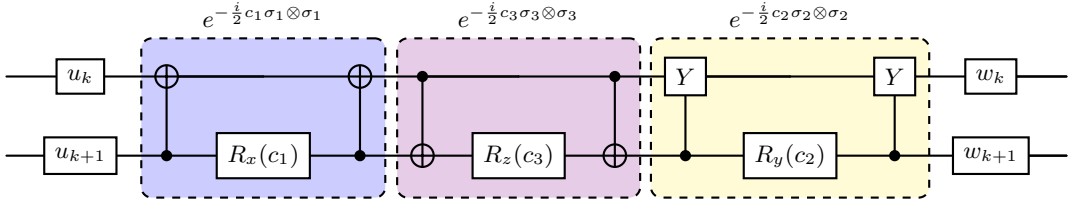
\begin{figure*}
\centering
\begin{quantikz}
	& \gate[1]{u_k} &  \targ{}\gategroup[2,steps=3,style={dashed,rounded corners,fill=blue!20, inner xsep=2pt},background]{$e^{-\frac{i}{2}c_1 \sigma_1\otimes \sigma_1}$} & \qw &\targ{} & \ctrl{1}\gategroup[2,steps=3,style={dashed,rounded corners,fill=violet!20, inner xsep=2pt},background]{$e^{-\frac{i}{2}c_3 \sigma_3\otimes \sigma_3}$} & \qw & \ctrl{1} & \gate{Y}\gategroup[2,steps=3,style={dashed,rounded corners,fill=yellow!20, inner xsep=2pt},background]{$e^{-\frac{i}{2}c_2 \sigma_2\otimes \sigma_2}$} & \qw & \gate{Y} &  \gate[1]{w_k} & \qw\\
	& \gate[1]{u_{k+1}} & \ctrl{-1} & \gate[1]{R_x(c_1)} & \ctrl{-1} & \targ{}& \gate[1]{R_z(c_3)} & \targ{} & \ctrl{-1} & \gate[1]{R_y(c_2)} & \ctrl{-1} & \gate[1]{w_{k+1}} & \qw
\end{quantikz}
\caption{Two-qubit unitary, $U$, at $k$ and $k+1$ th qubits. $U_{k,k+1} = (w_k\otimes w_{k+1})\mathcal{U}(u_k\otimes u_{k+1})$. $\mathcal{U}$ is shown in the Cartan form. $R_x(c_1), R_y(c_2), R_z(c_3)$ is the rotation about $X,Y,Z$ axes with angle $c_1, c_2, \text{ and }c_3$  respectively.}
\label{fig2}
\end{figure*}

At any given time, the two-qubit operator at sites $k$ and $k+1$ can be decomposed in the Cartan form as $U_{k,k+1} = (u_k\otimes u_{k+1})\, \mathcal{U}_{k,k+1}\, (w_k\otimes w_{k+1})$, where the interaction part $\mathcal{U}_{k,k+1}$ depends only on $3$ real parameters and characterizes the non-local content of $U_{k,k+1}$ \cite{ZhangVala2003, KrausCirac2001, Musz_2013}. We will choose $\mathcal{U}_{k,k+1}=\mathcal{U}$ to be independent of the site index $k$ as well as the time, {\em i.e.}, it remains the same throughout the circuit and is given by
\begin{equation}
\label{eq1}
\mathcal{U} =\exp{\left(-\frac{i}{2}\sum\limits_{j=1}^3 c_j\sigma_j^k \otimes \sigma_j^{k+1}\right)} 
\end{equation}
where, 
$\sigma_i$'s are the Pauli matrices. The three real parameters $\{c_1,c_2,c_3\}$ restricted such that $0\leq |c_3|\leq c_2 \leq c_1 \leq \pi/2$ comprise the tetrahedral Weyl chamber \cite{ZhangVala2003, Musz_2013}. This ensures that any two-qubit gate can be realized by sandwiching local unitary operators on $\mathcal{U}$, and that this parametrization is unique for any two locally equivalent unitaries. 
Figure~\ref{fig2} is the circuit representation of the Cartan decomposition of $U_{k,k+1}$.
The single qubit local unitaries, $u_k,u_{k+1},w_{k},w_{k+1}$, are allowed to vary not only along the sites $k$, but also across each time step, and are sampled from the Haar measure.


 Let $U$ be an unitary operator acting in a bipartite space $ {\cal H}_1\otimes {\cal H}_2$, $\mbox{dim}({\cal H}_{1,2})=d$. Its operator Schmidt decomposition,
$U= \sum_{i=1}^{d^2}\sqrt{\lambda_i} \, A_i \otimes B_i$,  
 where $A_i$ and $B_i$ are orthonormal operators on ${\cal H}_{1,2}$ respectively satisfying, $ \Tr(A_i^{\dagger}A_j)=\Tr(B_i^{\dagger}B_j)=\delta_{ij}$, $\lambda_i \geq 0$. As $U$ is unitary, $\sum_i \lambda_i = d^2$.
We  will consider operator entanglement as the linear entropy,
\begin{equation}
\label{eq:Elin}
E(U)= 1 - \sum_{i=1}^{d^2} \lambda_i^2/d^4.
\end{equation} 
This vanishes iff $U$ is a tensor product of two operators and takes the maximum value $E(S)=1-1/d^2$
where $S$ denotes  SWAP gate $S|\psi_1\rangle|\psi_2\rangle = |\psi_2 \rangle |\psi_1\rangle$.

Following \cite{Zanardi_2000}, the entangling power of a unitary operator is defined as
\begin{equation}
\label{eq:EntPower}
e_p(U)= \alpha \,  \overline{{\cal E}(|\psi \rangle = U|\psi_1\rangle |\psi_2\rangle)}^{|\psi_1\rangle, |\psi_2\rangle }
\end{equation}
with ${\cal E}$ being a suitable entanglement measure of {\it states}, and $\alpha$ is a constant that we will fix shortly. The average is taken over all the product states $|\psi_1\rangle$, $|\psi_2\rangle$
distributed uniformly. In this paper, we consider ${\cal E}$ to be  the linear entropy of the 
reduced density matrix $\rho_1=\Tr_{2}(|\psi\rangle \langle \psi|)$, that is ${\cal E}=1-\Tr(\rho^2_1)$.
 
It was shown in \cite{Zanardi_2000} that the entangling power $e_p(U)$ is simply related to the operator linear entanglement entropy $E(\cdot)$. Fixing the constant in the definition of the entangling power as $\alpha=(d+1)/(d-1)$, the following holds: 
 \begin{equation}
 e_p(U) = \frac{1}{E(S)} [E(U) + E(US) -E(S)],
 \end{equation}
 The constant is chosen such that the range of entangling power is independent of the local dimension $d$ and is $0\leq e_p(U)\leq 1$.

Local unitaries and the SWAP gate do not generate entanglement and hence have zero entangling power. Interestingly, this is despite SWAP being a non-local gate with the maximum operator entanglement $E(S)$. Gate typicality $g_t(U)$
originally defined in \cite{Jonnadula_2017, Jonnadula_2020}, distinguishes local and non-local gates and is given by, 
\begin{equation}
 g_t(U) = \frac{1}{2E(S)} [E(U) - E(US) + E(S)]. 
 \end{equation}
 The range of gate typicality is $0\leq g_t(U)\leq 1$ \cite{Jonnadula_2017, Jonnadula_2020}, and is maximum only for the SWAP gate and operators that are locally equivalent to it, $g_t(S)=1$.
Qualitatively, gate typicality captures the amount of non-local resources needed to construct a gate. For example, one needs three CNOT gates to construct a single SWAP gate and two CNOT is needed for DCNOT which is SWAP~$\times $ ~CNOT. The corresponding values of $g_t$ are $g_t(\text{CNOT})=1/3$, $g_t(\text{DCNOT})=2/3$. However, an interpretation of gate typicality, similar to entangling power, is still lacking. In this work, we will show the importance of gate-typicality in MIPT in determining the criticality.

All of the quantities $E(U)$, $E(US)$, $e_p(U)$ and $g_t(U)$ as defined above are LU invariants, any two of them being linearly related to the other two.
In terms of the Cartan coefficients of Eq.~(\ref{eq1}) we have \cite{Jonnadula_2020},
\begin{equation}
\label{eq:epgt_formulas}
\begin{split}
e_p(U) =& \frac{2}{3}\left( \sin^2{c_1} \cos^2{c_2} + \sin^2{c_2} \cos^2{c_3} + \sin^2{c_3}\cos^2{c_1} \right)\\
g_t(U) =& \frac{1}{3} \left( \sin^2{c_1} + \sin^2{c_2} + \sin^2{c_3}\right),
\end{split}
\end{equation}

\begin{figure}[!htbp]
\centering
\includegraphics[width=0.8\linewidth]{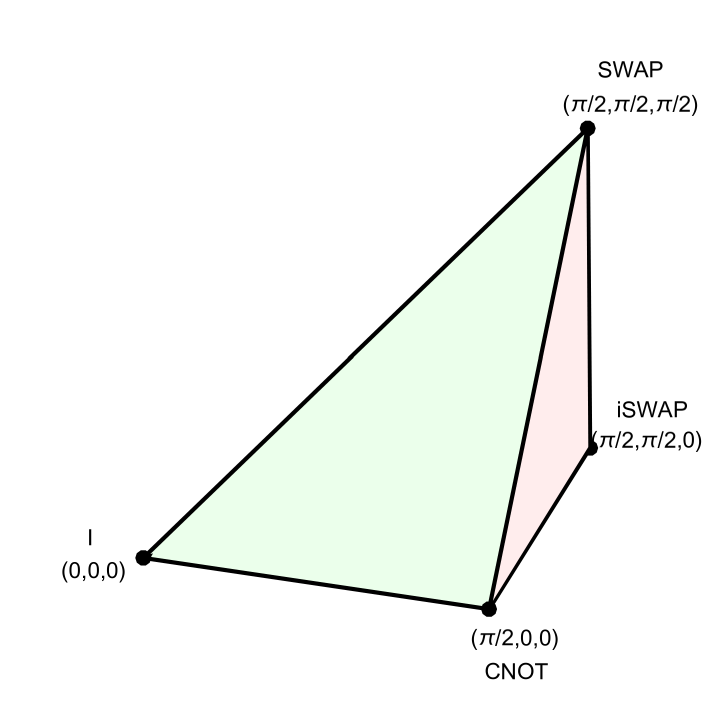}
\caption[]{Each two-qubit gate is uniquely determined up to single-qubit unitaries by three parameters restricted such that $0\leq |c_3|\leq c_2 \leq c_1 \leq \pi/2$  that form the Weyl chamber. The gates used in this paper are sampled from the two faces of the tetrahedron shown here. }
\label{weyl_faces}
\end{figure}

\begin{figure}[!htbp]
\centering
\includegraphics[width=0.8\linewidth]{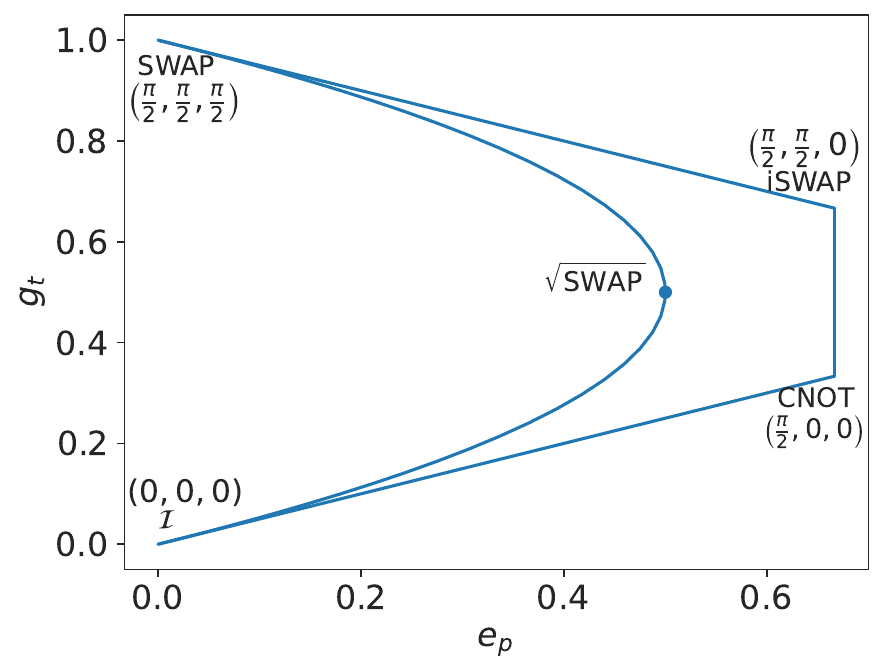}
\caption[]{Two-qubit gates represented by entangling power and gate typicality in the $e_p-g_t$ plane. The identity, $\mathcal{I}$, SWAP, iSWAP (locally equivalent to DCNOT), and CNOT are the four vertices that mark the boundary of the set of permissible two-qubit gates. Also shown are the Cartan coefficients $\{c_1,c_2,c_3\}$ for the four vertices.}
\label{fig3}
\end{figure}

Every two-qubit gate $U$ can be represented as a point in a two-dimensional plane, the ``$e_p$-$g_t$ plane'', with the entangling power $e_p$ and gate typicality $g_t$ as the axes. Since both entangling power and gate typicality are LU invariant, all LU counterparts of $U$ are represented by the same point. 
Moreover, every two-qubit gate lies in a bounded region of the $e_p$-$g_t$ plane  \cite{Jonnadula_2020}, as shown in  Fig.~\ref{fig3}. In our analysis, we consider a set of $\mathcal{U}$ that lie on two faces of the Weyl chamber as shown in Fig.~\ref{weyl_faces}. This set covers the full $e_p$-$g_t$ plane of  Fig.~\ref{fig3}. The identity $\mathcal{I}$ and SWAP gates located at the left vertices have zero entangling power. The other two vertices, the CNOT and the DCNOT (locally equivalent to iSWAP) have maximal entangling power. 
As one traverses the boundary in Fig.~\ref{fig3}, along the $\mathcal{I} - \text{CNOT}$ and  $\text{SWAP} -\text{iSWAP}$ lines, the entangling power monotonically increases.  

The gates that lie on the $\text{SWAP} -\text{iSWAP}$ line, are such that their operator entanglement $E(U)$ is maximal, that is $E(U)=E(S)$. These are identical to  ``dual-unitary" operators which enjoy a ``space-time duality" and circuits constructed from these are solvable in some ways \cite{BertiniKosProsen, AravindaSuhail2021}.
Those on the $\mathcal{I} - \text{CNOT}$ line have been called ``T-dual unitary'' \cite{AravindaSuhail2021}, as their partial transposes are also unitary and they are such that $E(US)=E(S)$ is maximized, in other words they are products of dual unitaries with the SWAP operator. Thus the choice of these gates picks out a large set of interesting ones, with special properties.

The Cartan form of Eq.~\eqref{eq1} shows that there are three invariants, the $c_j$ of which the entangling power and gate-typicality are functions of. Therefore there is a third invariant in general. However, for the gates considered in the paper, that lie on the planes in Fig.~\ref{weyl_faces}, two invariants are sufficient. In other words for this set of gates, their representation in the $e_p$-$g_t$ plane is unique. A choice of the third invariant and its interpretation is interesting and explorations of gates in the interior of the Weyl chamber should reveal its possible role in determining the MIPT, however we leave this for future work. 

The parabolic curve $\mathcal{I} - \text{SWAP}$, is given by the fractional powers of the SWAP operator \cite{Jonnadula_2020}. Lastly, all gates have the maximum possible entangling power ($=2/3$) along the vertical $\text{CNOT}$ to $\text{iSWAP}$ line with a monotonically increasing gate typicality. The fact that for two-qubit gates the entangling power does not reach $1$, is due to the non-existence of gates that are simultaneously dual and T-dual. Such gates exist for all local dimension $>2$ and are examples of 2-unitaries or perfect tensors \cite{DardoKarol2015,SuhailPRl2022}.

Returning to the many-body scenario, in the quantum circuit in Fig.~\ref{fig1}, while the two-qubit operators $U_{k,k+1}$ are all different, their Cartan decomposition's nonlocal part $\mathcal{U}$ are identical throughout the circuit and hence $e_p(U_{k,k+1})=e_p(\mathcal{U})$ and similarly for $g_t(U)$. After a finite number of time steps involving the unitary and measurement layers, we find the entanglement entropy of the first half of the circuit with the other. 
Our focus is to study the transition of entanglement from a volume to an area law as we increase the probability of measurement in the circuit. Therefore, we pick our initial state of the $L$ qubits from the Haar measure of random states in the $2^L$ dimensional space.
Thus our initial states are already scrambled and have near-maximal bipartite entanglement given by the Page value $L(\ln{2}/2)+ \text{constant}$ \cite{Page_1993}.  Repeated measurements will seek to destroy this large entanglement, while the unitary evolutions denoted generically by $V$ will seek to preserve or enhance.
We investigate the entanglement entropy at a sufficiently long time ($t=2L$) as a function of the measurement probability $p$. By averaging over multiple circuit realizations with the same $\mathcal{U}$, we establish connections between the measurement-induced phase transition (MIPT) and both $e_p(\mathcal{U})$ and $g_t(\mathcal{U})$.

\section{Entanglement across a hybrid circuit}
\label{change of entropy}
In this section, we address one of the central questions that is the focus of our study. We explore the role played by entangling power and gate typicality of the two-qubit unitaries of the hybrid circuit in the amount of global bipartite entanglement sustained by the circuit. For this purpose we evolve a circuit with $L$ qubits for 2$L$ time steps and apply random measurements with a given probability $p$ after each time step.

\subsection{{Measurement only circuit}}
When the non-local part of the circuit, $\mathcal{U}$, is the identity, the behavior of entanglement is only governed by measurements.
In this case, we can analytically estimate the entanglement $\mathcal{S}$ as a function of measurement probability $p$. In this scenario as soon as a qubit is measured, it will disentangle from the rest of the qubits permanently. If we apply the measurement step once with probability $p$, the probability that any single qubit will not be measured is $(1-p)$. After $t$ time steps the probability that this qubit is not measured is $(1-p)^t$. We will measure the entanglement entropy between part $A$, consisting of the first $N=L/2$ qubits, and part $B$, comprising of the remaining $L/2$ qubits. After $t$ time step probability that $N_A$ qubits in part $A$ are unmeasured is,
\begin{equation}
    \label{partaAprob}
    p_A(N_A,p,t)= \binom{N}{N_A}(1-p)^{tN_A} \left(1-(1-p)^t \right)^{N-N_A}.
\end{equation}
Similarly, the probability that $N_B$ qubits remain unmeasured in part $B$, $p_B(N_B,p,t)$, is given by 
the same expression as $p_A$ except that $N_A$ is replaced by $N_B$.
Since each qubit in the circuit is measured independently of the others, the probability that $N_A$ qubits in part $A$ and $N_B$ qubits in part $B$ remain unmeasured is $p_A(N_A,p,t)\cdot p_B(N_B,p,t)$. If $S(N_A:N_B)$ denotes the bipartite entanglement entropy, then the expected entanglement entropy $\mathcal{S}$ is
\begin{equation}
    \label{onlyMS1}
    \mathcal{S}(p,t) = \sum_{N_A,N_B=0}^N  S(N_A:N_B) p_A(N_A,p,t) p_B(N_B,p,t).
\end{equation}
The entanglement entropy of a Haar random state follows the Page value \cite{Page_1993}. In order to obtain an analytical estimate on the expected entropy, we assume that if $x$ qubits are measured in a Haar random state, the resulting state has an entanglement entropy equivalent to that of a Haar random state with the remaining $L-x$ qubits. Based on this assumption, we can express the bipartite entanglement between parts $A$ and $B$ as
\begin{equation}
    \label{page_value}
    S(N_A:N_B) = \sum_{k=2^{\max(N_A,N_B)}}^{2^{N_A+N_B}}\frac{1}{k} - \frac{2^{\min(N_A,N_B)}-1}{2^{\max(N_A,N_B)+1}}.
\end{equation}

\begin{figure}[!htbp]
\centering
\includegraphics[width=0.9\linewidth]{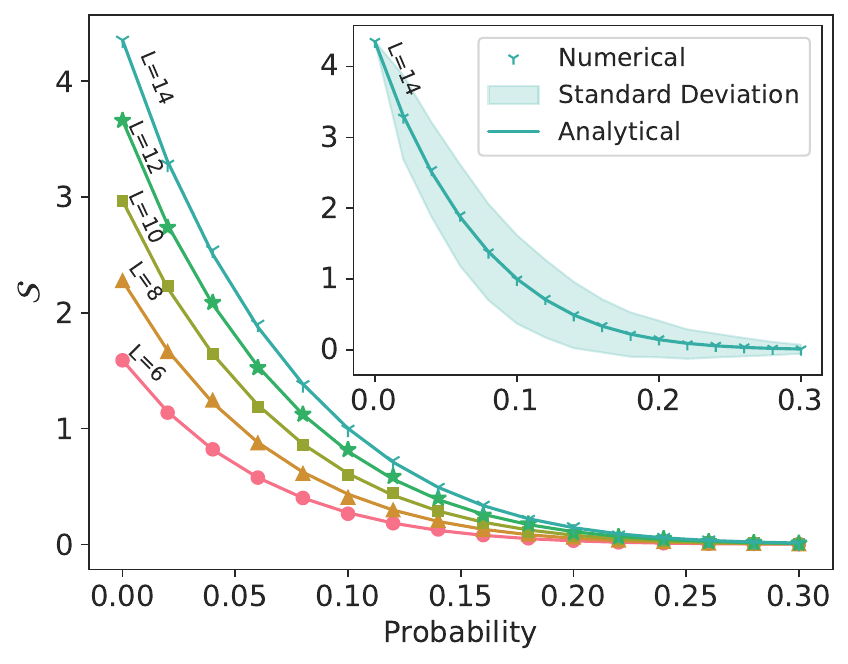}
\caption[]{Entanglement entropy with probability of $\mathcal{U}= \mathcal{I}$. In the plots, points with different markers correspond to the average entanglement obtained by simulating 1500 instances of the circuit with a given probability of measurement over 10 time steps. The solid lines are from the theoretical estimate given by Eq.~(\ref{onlyM_final}). Inset shows entanglement entropy curves plotted against probability for $L=14$ along with the standard deviation.}
\label{fig:Monlyth}
\end{figure}

\begin{figure}[!htbp]
\centering
\includegraphics[width=0.9\linewidth]{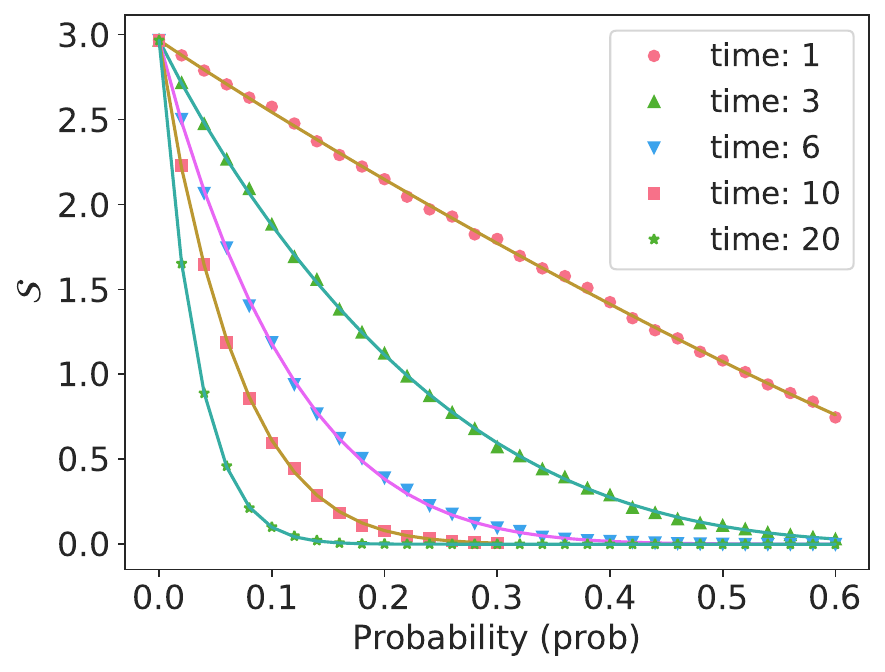}
\caption[]{Bipartite entanglement entropy with probability for $\mathcal{U} = \mathcal{I}$ of 10 qubits at different time steps. In the plot points with different markers correspond to the average entanglement obtained by 1500 instances of the circuit with a given probability of measurement. Solid lines are the analytical results from Eq.~\eqref{onlyM_final}.} 
\label{fig:Monlyth_dtime}
\end{figure}

Combining Eq.~(\ref{partaAprob},\ref{onlyMS1},\ref{page_value}), we obtain an estimate for the expected entropy as 
\begin{widetext}
	\begin{equation}
		\label{onlyM_final}
		\mathcal{S}(p,t) = \sum_{N_A,N_B=0}^N  S(N_A:N_B) \binom{N}{N_A} \binom{N}{N_B} (1-p)^{t(N_A+N_B)}  \left(1-(1-p)^t \right)^{2N-(N_A+N_B)}.
	\end{equation}
\end{widetext}

We find a remarkable agreement between our numerical results and the analytical expression derived above. In Fig.~\ref{fig:Monlyth} different markers denote the average entanglement entropy computed from simulating 1500 instances of the circuit for various measurement probabilities over 10 time steps.
The solid line represents the theoretical prediction according to Eq.~(\ref{onlyM_final}). 
The Fig.~\ref{fig:Monlyth_dtime}, displaying entanglement entropy curves against probability for 5 different time steps $t$ utilizing 
$L=10$ qubits, also shows an excellent agreement with the theoretical estimate. 

\subsection{Entanglement along the boundary}
We first calculate the entanglement entropy as a function of the measurement probability along the boundaries of Fig.~\ref{fig3}. Our results are shown in  Figs.~\ref{fig:fracSWAP}, \ref{fig:fraciSWAP}, \ref{fig:fracCNOT}, \ref{fig:fracCNOTiSWAP}. Each point of the plot represents the average  entanglement over many circuit iterations after $2L$ time steps. For each circuit iteration, we start with a different Haar random state as the input while keeping the same unitary $\mathcal{U}$ and the same measurement probability. 

\subsubsection{Fractional powers of SWAP}
\begin{figure}[!htbp]
\centering
\includegraphics[width=0.9\linewidth]{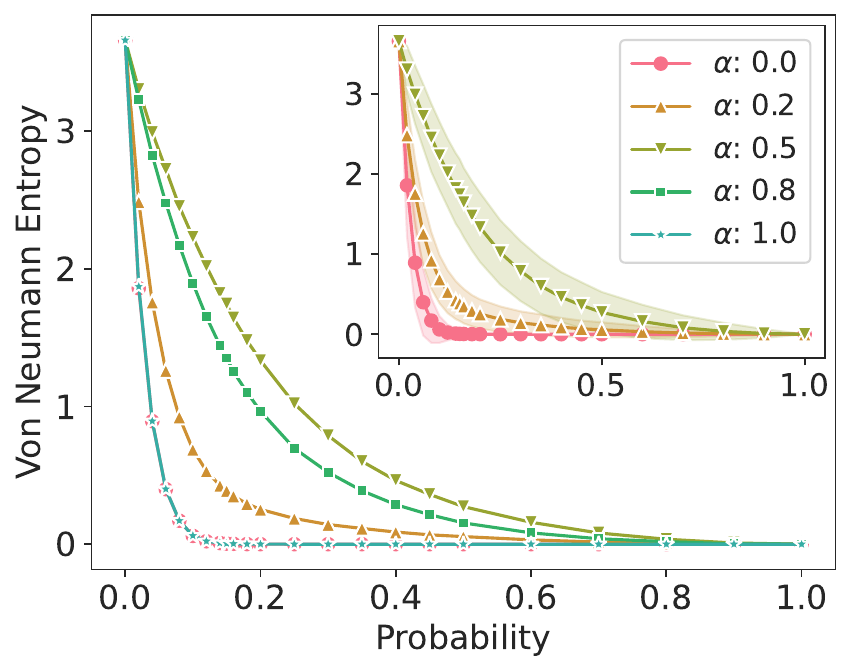}
\caption[]{Decay of entanglement entropy as a function of probability of measurement along the parabolic boundary of Fig.~\ref{fig3} for $L=12$. Five curves show different gates along this line and can be expressed as $(\text{SWAP})^\alpha$. Here, $\alpha \in [0,1]$, $\alpha = 0$ corresponds to the identity, $\mathcal{I}$,  and $\alpha=1$ is the SWAP  gate.  Inset shows the same plot with standard deviation (filled region) corresponding to $\alpha$ values: $0,\, 0.2,$ and $ 0.5$.}
\label{fig:fracSWAP}
\end{figure}

In Fig.~\ref{fig:fracSWAP}, we choose the two-qubit gates as $\mathcal{U}=S^{\alpha}$ along the $\mathcal{I} -\text{SWAP}$ curve of Fig.~\ref{fig3}. Gates along this parabolic boundary are parameterized by $c_1 =c_2=c_3= \alpha \pi/2$, with  $\alpha \in [0,1]$. The end points of the interval corresponding to the identity $\mathcal{I}$ and SWAP gate respectively. 
This curve can be viewed as consisting of two parts. From $\mathcal{I} -\sqrt{\text{SWAP}}$ the entangling power $e_p(\mathcal{U})$ increases with the maximum value being attained at $\sqrt{\text{SWAP}}$ ($\alpha = 0.5$). 

The drop in the entanglement of the hybrid circuit constructed from such an $\mathcal{U}$ and random local unitaries is sharper for those $\mathcal{U}$ with smaller entangling power as one increases measurement probability. This can be seen in Fig.~\ref{fig:fracSWAP} with
$\mathcal{I}$ having the sharpest drop in the entanglement entropy.  
From $\sqrt{\text{SWAP}} -{\text{SWAP}}$ along the same curve, $e_p({\mathcal{U}})$ decreases and once again we observe a sharper fall in entanglement values. This makes sense as the unitary dynamics increases the entanglement in the circuit while projective measurements tend to break entanglement. This trade off makes entanglement, for a given measurement probability, settle at a higher value for unitaries with greater entangling power.
Another interesting observation from Fig.~\ref{fig:fracSWAP} is that the entanglement entropy curves for $\mathcal{I}$ and $\text{SWAP}$ overlap. The entangling power for both is zero and they show the fastest entropy drop with an increase in measurement probability. In contrast, $\sqrt{\text{SWAP}}$ has the maximum entangling power along this curve and shows a most gentle drop in the value of entanglement entropy with an increase in measurement probability. Although the gates $\mathcal{U=}S^\alpha$ and $S^{1-\alpha}$ have the same entangling power, the  fall of entropy as a function of measurement probability in hybrid circuits built with them with are significantly different. This is seen in the Fig.~\ref{fig:fracSWAP} in the cases of $\alpha=0.2$ and $\alpha=0.8$, indicating a role for a quantity such as the gate typicality.

\subsubsection{Dual unitary gates (SWAP to iSWAP)}

\begin{figure}[!htbp]
\centering
\includegraphics[width=0.9\linewidth]{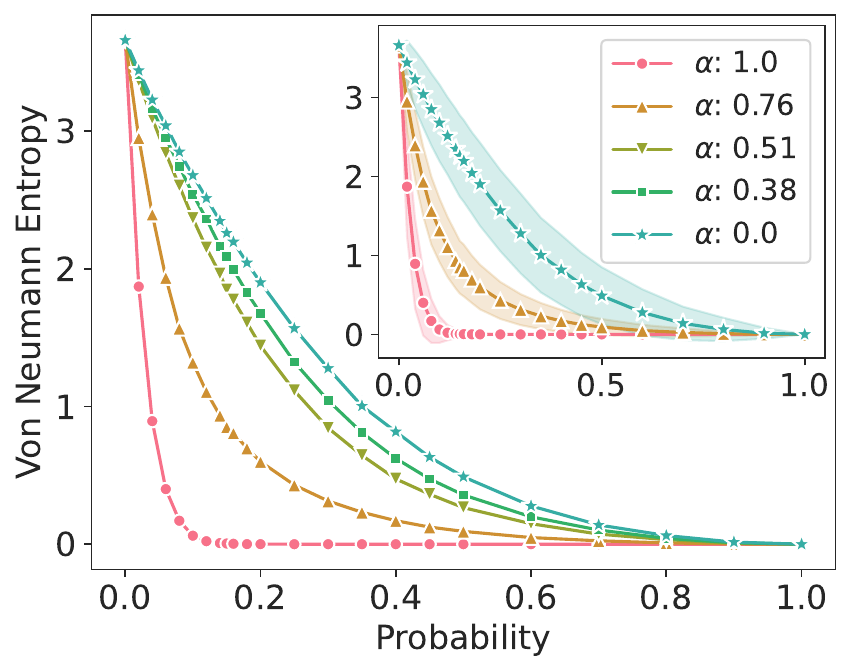}
\caption[]{Decay of entanglement entropy as a function of probability of measurement along the $\text{SWAP}-\text{iSWAP}$ line of Fig.~\ref{fig3} for $L=12$.
Five curves shown are different gates along this line and can be expressed as $(\text{SWAP})^\alpha ( \text{iSWAP})^{1-\alpha}$ with $\alpha \in [0,1]$. Inset shows the same plot with standard deviation (filled region) for $\alpha$ values:  $0,\, 0.76,$ and $ 1.0$.}
\label{fig:fraciSWAP}
\end{figure}

Figure~\ref{fig:fraciSWAP} shows the entanglement entropy as a function of the measurement probability for the circuit as the interaction  $\mathcal{U}$ is varied  along the SWAP to iSWAP (or DCNOT) line of Fig.~\ref{fig3}. 
These gates parameterized by the Cartan coefficient $c_3$ while keeping $c_1$ and $c_2$ fixed at $\pi/2$, maximize operator entanglement, $E(\mathcal{U})=E(S)$. Thus they are ``dual-unitary", and as $c_3$ varies from $\pi/2$ to $0$ where $\pi/2$ and $0$ correspond to the SWAP and iSWAP gates respectively, the entangling power changes from $0$ to the maximum possible value of $e_p(\text{iSWAP})=2/3$. The interaction can equivalently be written simply as  $\mathcal{U} = (\text{SWAP})^\alpha ( \text{iSWAP})^{1-\alpha}$, where $c_3=\pi \alpha/2$.

The variation observed for the different $\mathcal{U}$ aligns with the picture above. 
We observe that hybrid circuit has a sharper fall of entanglement for smaller $e_p(\mathcal{U})$. The circuit constructed from the SWAP gate ($\alpha=1$) with zero entangling power shows the fastest fall of entanglement with probability. The iSWAP gate displays the slowest fall of entanglement with probability. The intermediate cases show a monotonic correlation between the entangling power of $\mathcal{U}$ and the entanglement present in the hybrid circuit.

\subsubsection{T-dual gates ($\mathcal{I}$ to CNOT) }

\begin{figure}[!htbp]
\centering
\includegraphics[width=0.9\linewidth]{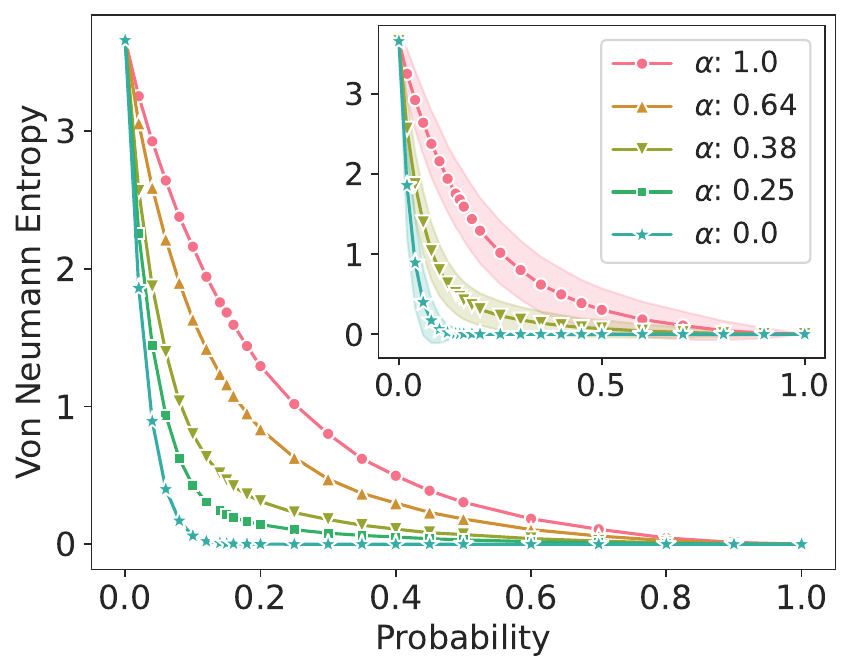}
\caption[]{Decay of entanglement entropy as a function of  probability of measurement along the $\mathcal{I}- \text{CNOT}$ line of Fig.~\ref{fig3} for $L=12$. Five curves show different gates along this line and can be expressed as $(\text{CNOT})^\alpha$.
Here $\alpha = 0,1$ corresponds to the Identity $\mathcal{I}$ and the CNOT gate respectively. Inset shows the same plot with standard deviation (filled region) corresponding to $\alpha$ values: $0,\, 0.38,$ and $ 1.0$.}

\label{fig:fracCNOT}
\end{figure}

In Fig.~\ref{fig:fracCNOT}, we traverse along the  $\mathcal{I}$ to CNOT line, taking for $\mathcal{U}$ five unitaries of increasing entangling power and finding exactly similar behavior as discussed above. All the gates along this line can be expressed as $(\text{CNOT})^\alpha$ where $\alpha \in [0,1]$. Here $\alpha = 0,1$ corresponds to the Identity $\mathcal{I}$ and the CNOT gate respectively.
In the Cartan form, these gates are  parameterized by $c_1$, keeping  $c_2$ and $c_3$ fixed at $0$. We change $c_1$ from $0$ to $\pi/2$  with $0$ and $\pi/2$  corresponding to $\mathcal{I}$ and CNOT respectively.
Unitaries with smaller entangling power have a sharper drop of entanglement with increasing measurement probability. In both Fig.~\ref{fig:fraciSWAP} and Fig.~\ref{fig:fracCNOT}, the curves for the two-qubit unitaries with a lower entangling power lie below those with a higher value.

\subsubsection{ $\mathcal{U}$ from the CNOT to iSWAP line}

\begin{figure}[!htbp]
\centering
\includegraphics[width=0.9\linewidth]{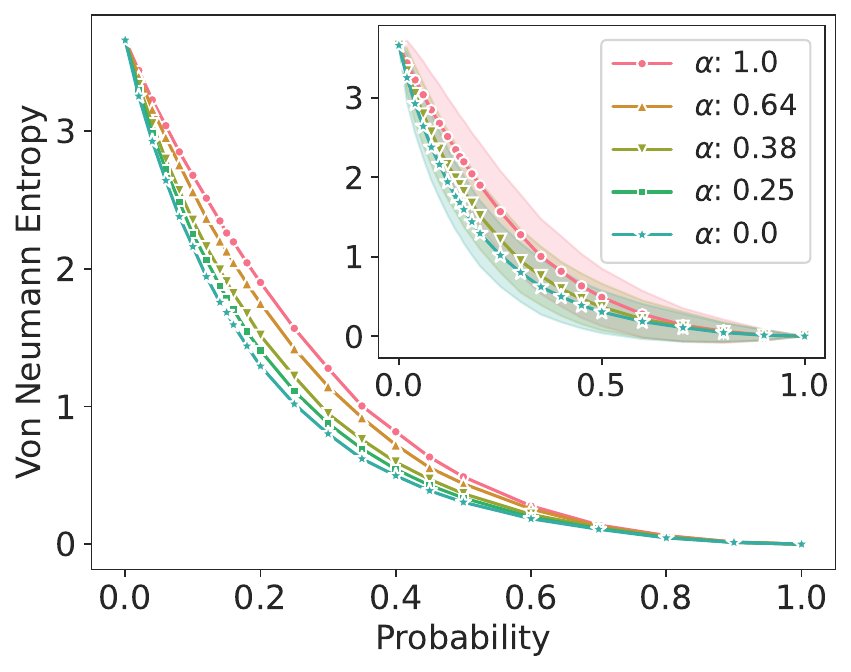}
\caption[]{Decay of entanglement entropy as a function of probability of measurement along the CNOT to iSWAP boundary of Fig.~\ref{fig3} for $L=12$. All the gates along this line can be written as $(\text{CNOT})^{1-\alpha}(\text{iSWAP})^\alpha$. Inset shows the same plot with standard deviation (filled region) corresponding to $\alpha$ values: $0, 0.38,$ and $ 1.0$.}
\label{fig:fracCNOTiSWAP}
\end{figure}

It is remarkable that we are able to correlate the behavior of entanglement entropy for a large class of unitaries by continuously varying the entangling power, a single parameter, and give an intuitive explanation. However, as already indicated in the case of $\mathcal{U}$ being fractional powers of SWAP, the entangling power may not be the complete story. 
Figure~\ref{fig:fracCNOTiSWAP} confirms this as along the line CNOT to iSWAP (or DCNOT) all gates have the maximum possible entangling power ($=2/3$). Gates on this line can be written as $(\text{CNOT})^{1-\alpha}(\text{iSWAP})^\alpha$, and are parameterized by the Cartan coefficient $c_2=\pi \alpha/2$ while keeping $c_1=\pi/2$ and $c_3=0$ fixed. As we change $c_2$ from $0$ to $\pi/2$, we move from CNOT to iSWAP and the gate typicality monotonically increases. Figure~\ref{fig:fracCNOTiSWAP} shows a monotonic increase in entanglement entropy for a fixed measurement probability along the CNOT-iSWAP line.  Therefore, gate typicality plays a important role in determining the entanglement entropy in hybrid circuits. 


\begin{figure}[!htbp]
\centering
\includegraphics[width=0.9\linewidth]{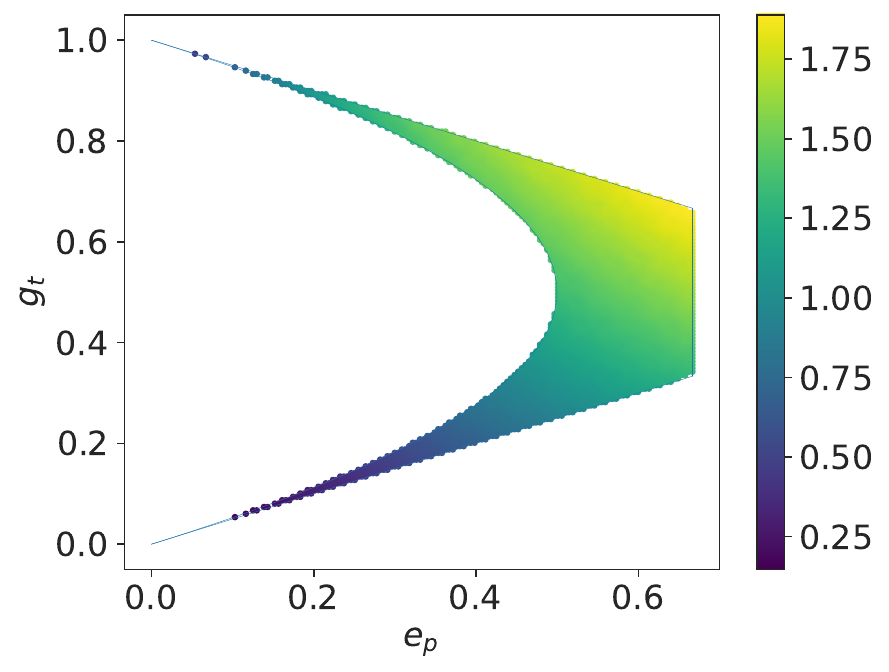}
\caption[]{Entanglement entropies for the entire spectrum of non-local unitaries in the $e_p-g_t$ plane with the measurement probability $p = 0.2$ for 12 qubits. The color bar indicates the entanglement between two halves of the system. The  ``heat-map" is obtained by uniformly sampling 614 unitaries from the interior of the permissible region in the $e_p-g_t$ plane and calculating the entanglement entropy for $p=0.2$ and smoothing out the surface by interpolation.}
\label{fig5}
\end{figure}

This claim is further strengthened when $\mathcal{U}$ is taken from unitaries that lie inside the boundaries.Figure~\ref{fig5} shows the ``heat-map'' of the entanglement entropy sustained in hybrid circuits, as a function of the $e_p-g_t$ values of the interaction unitary $\mathcal{U}$, when the measurement probability is $p=0.2$.
Consider a line parallel to the vertical axis of this figure. All unitaries along this line lying in the interior region have a fixed entangling power with a monotonically increasing value of gate typicality. We observe, for a fixed measurement probability, an increase in entanglement entropy along this line. Therefore, the amount of bipartite entanglement across the entire hybrid circuit can be ordered by gate typicality of the constituent unitary components when they have a fixed entangling power. Our work also provides another operational interpretation for gate typicality. Gate typicality serves as an unambiguous criterion to order a family unitaries, with a fixed entangling power, by the amount of global bipartite entanglement they can generate.

 Traversing across a horizontal line towards the positive x-axis, we see that entanglement entropy is positively correlated with the entangling power. 
From Fig.~\ref{fig5} it is clear the maximum entanglement sustained is in the north-east corner of the plot at the iSWAP vertex. Not surprisingly, the south-west corner at the $\mathcal{I}$ vertex is the region of minimum possible entanglement that can be sustained in the hybrid circuit.


\section{MIPT as a function of two-qubit gate properties}
\label{phase_transition}
The hybrid circuit that we have considered is a toy model for a many-body system with interactions. It has been shown that, when measurements occur randomly with a probability $p$ and the unitary gates employed are Haar random,  the system has two dynamical phases. One phase characterized by volume-law entanglement occurs below a critical threshold value of $p$, i.e., for $p<p_c$. Above this threshold, for $p>p_c$, the system is characterized by an area-law entanglement. At $p=p_c$, there is a phase transition. 

In a system with $d$ spatial dimensions, consider a subsystem $A$ with volume $|A|^d$. If the entanglement entropy of this subsystem scales as $\mathcal{S} \sim |A|^d$, it is said to follow the volume law of entanglement. In one spatial dimension, this indicates a direct proportionality to the subsystem size ($l$). However, additional regimes of entanglement exist. For example, when the system resides in a ground state or in a localized phase, as discussed in previous studies\cite{RevModPhys.82.277, PhysRevLett.119.110604}, the entanglement follows a different scaling correlating with the boundary of the subsystem size. This phenomenon, known as the area law of entanglement, is characterized by $\mathcal{S} \sim |A|^{(d-1)}$, indicating short-range entanglement. In the specific case of a one-dimensional system ($d=1$) the volume law phase corresponds to $\mathcal{S} \sim l$ and the $\mathcal{S}$ under area law remains independent of the system size. Considering a bipartite system with Hilbert space dimensions, $d_A$ and $d_B$, with $d_A \leq d_B$, the average entanglement entropy of a typical random state can be expressed as $\mathcal{S} \approx \ln{d_A} - \frac{d_A}{2d_B}$\cite{Page_1993}. For our scenario assuming two equal partitions of $L$ qubits, 
we have $d_A = d_B = 2^{L/2}$. 

Given that our initial state is a Haar random state and remains so without measurement, the entanglement entropy shows a linear scaling with the system size and is given by $\mathcal{S} = \frac{L}{2}\ln{2} - 0.5$. We may expect the same size scaling for small values of $p$.
For the area law, entanglement entropy is a constant, independent of $L$, that depends only on $p$. 
For the three different situations, entanglement entropy is \cite{Li_2019}

\begin{equation}
\label{S1}
\mathcal{S} = \left\{ \begin{array}{rcl}
a(p)L+b(p) \ln{L}  & \mbox{at}
& p< p_c \\ a(p_c)\ln{L}
& \mbox{at} & p=p_c \\
c(p) & \mbox{at} & p> p_c
\end{array}\right.
\end{equation}
where $a(p)$, $b(p)$ and $c(p)$ are the functions of $p$.


The goal of this work is to extend the study of measurement-induced phase transitions beyond the case of two-qubit Haar random gates to any two-qubit unitary gate parameterized by entangling power and gate typicality. 
 To start with, we consider the application of CNOT gates as the non-local unitary to be applied throughout the hybrid circuit. As discussed above, interfacing local Haar random single unitary gates with the CNOT ensures we are considering all the locally equivalent versions of the CNOT.  
 This family of gates has the maximal entangling power and, as shown in Fig.~\ref{fig3}, lies on one of the vertices. 
 
\begin{figure}[!htbp]
\centering
\includegraphics[width=0.9\linewidth]{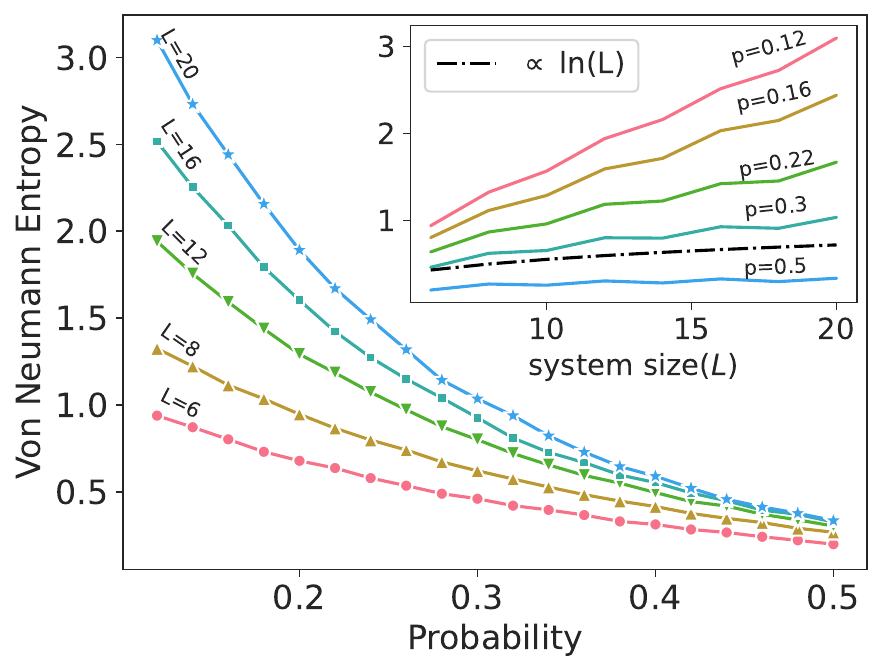}
\caption[]{The von Neumann entropy as a function of probability of measurement. Inset figure is the variation of entanglement entropy with system size for different values of measurement probabilities. Observable regions include the volume law (proportional to system size), area law (remaining relatively constant), and a critical point (p = 0.37, scaling as $\ln{L}$). Three regions of entanglement are shown - volume law $(\sim L)$ for $p< 0.37$, area law $(\sim constant)$ for $p > 0.37$ and the critical point (indicated by a red line at $p=0.37$, scaling as $\sim \ln{L}$).}
\label{fig6}
\end{figure}


The behavior of entanglement entropy of the many-body circuit versus the measurement probability for varying system sizes is shown in Fig.~\ref{fig6}. 
As observed in previous MIPT studies \cite{Koh_2023, Skinner_2019}, there is a montonic decrease of the entanglement entropy with the probability of measurements, till a putative area-law is reached. This is borne out more clearly
in the inset, where the entanglement entropy versus system size is plotted for different measurement probabilities.  Even for system sizes of a few qubits ($6\leq {L}\leq 20$), signatures of what would be a phase transition in the thermodynamic limit can be seen, as between $p=0.5$ and $p=0.3$, a critical growth of entanglement  $\sim \ln{L}$ occurs. 

Rigorous numerical analysis using scaling functions \cite{Calabrese_2005, Skinner_2019} and numerical optimization, discussed below, shows a phase transition at $p_c = 0.37$
with a logarithmic growth of entanglement consistent with equation Eq.~\ref{S1}.
In particular a departure from the volume-law entanglement, for $p<p_c$ to an area law for $p>p_c$ approximately around $p_c = 0.37$ can be observed.

\begin{figure}[!htbp]
\centering
\includegraphics[width=0.9\linewidth]{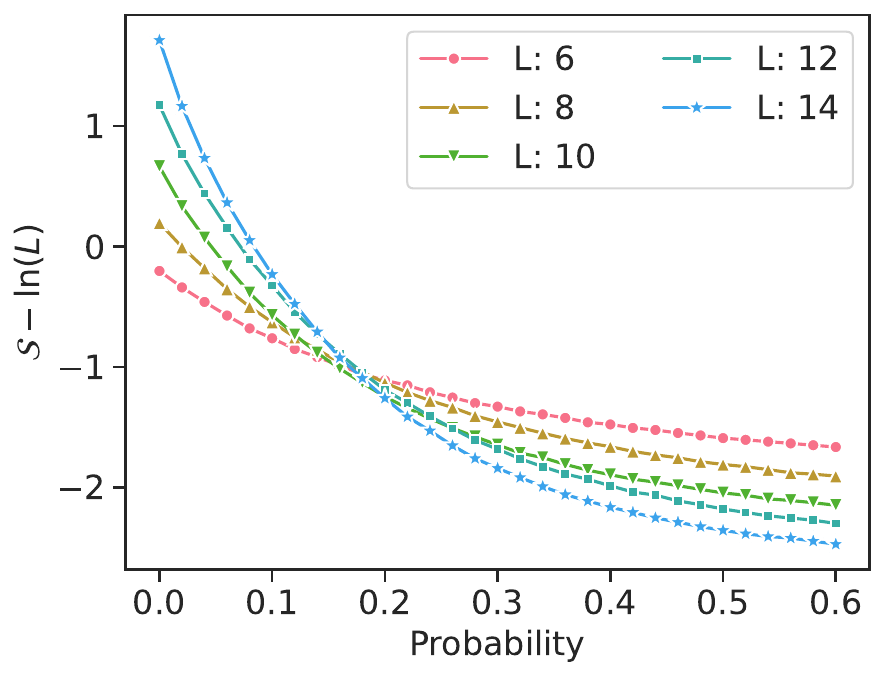}
\caption[]{Relationship between rescaled entropy and the probability of measurement, revealing the presence of a phase transition for CNOT as a non-local gate.}
\label{fig8}
\end{figure}

For systems of the order of a few qubits, we can get an estimate of critical probability $p_c$ by  simply shifting the entropy by $\ln{L}$. Although this only gives an approximate value of the critical probability, much like our analysis leading to Fig.~\ref{fig6}, it is worth describing this for its intuitive appeal.  
 Figure~\ref{fig8} shows the plot of $\mathcal{S} -\ln{L}$ {\em vs} $p$ and one can confirm these observations. The shifted entanglement entropy, for varying system sizes, tends to cross in a small band of probability values, indicating a MIPT.
 \begin{figure}[h!]
\centering
\includegraphics[width=1.0\linewidth]{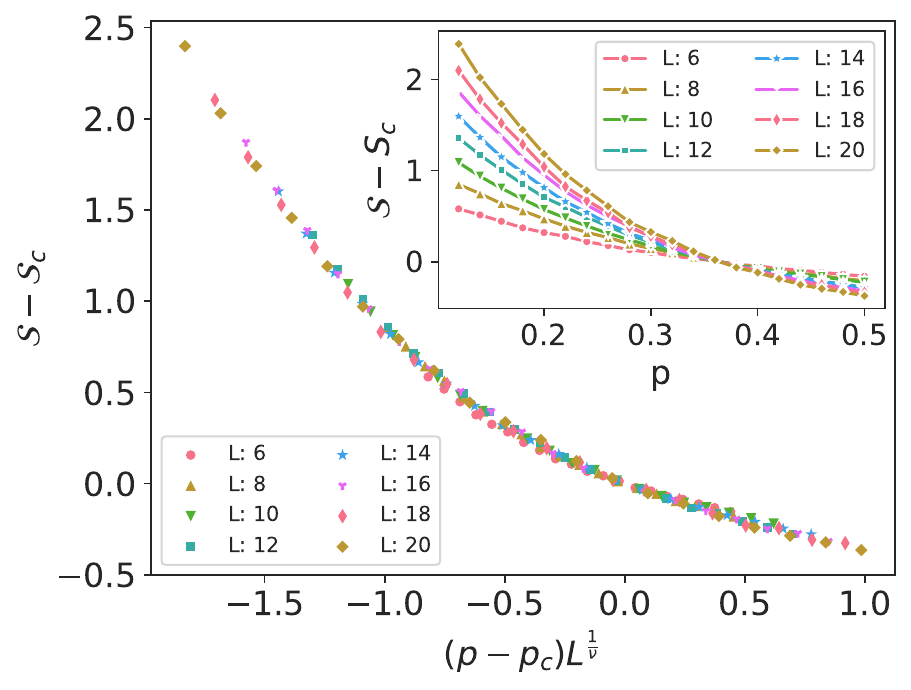}
\caption[]{Data collapse for CNOT using eight distinct system size to determine the critical probability, $p_c$, and the critical exponent, $\nu$.  $p_c = 0.367 \pm 0.007$ and $\nu = 1.495 \pm 0.095$}
\label{fig9}
\end{figure}

The exact value of $p_c$ is found using finite size scaling and data collapse. Here one starts by choosing an appropriate scaling function with parameters whose best fit gives the critical value of the probability.  Using a similar analysis as an earlier work  on MIPT \cite{Skinner_2019}, the scaling function $F$ can be expressed as 
\begin{equation}
    \label{eq6}
    F[(p-p_c)L^\frac{1}{\nu}] = \mathcal{S}-\mathcal{S}_c.
\end{equation}

In the $\mathcal{S} - \mathcal{S}_c$ {\em vs} $(p-p_c)L^\frac{1}{\nu}$ plot, as shown in Fig.~\ref{fig9} for the case of $\mathcal{U}=\text{CNOT}$, we find the best fit $p_c$ and $\nu$ so that curves for different systems sizes fall on each other. All curves for entanglement entropy with varying system sizes collapse on a single curve and the origin will give the exact value of $p_c$, the critical probability. We find that the value of $p_c$ for phase transition in hybrid circuits with CNOT as the non-local gate to be $ 0.367 \pm 0.007$. Furthermore, the critical exponent obtained is $\nu = 1.495 \pm 0.095$.  We note that for random unitary circuits \cite{Skinner_2019} the MIPT critical point, $p_c$ and $\nu=2$, is very different from the one obtained by fixing the interacting gate as CNOT.


\section{Critical probability with entangling power and gate typicality}
\label{pc_ep_gt}

In our investigation detailed in Section \ref{change of entropy}, we unraveled an intriguing relationship between entanglement entropy across a hybrid circuit on the one hand, and entangling power and gate typicality of two-qubit unitary gates on the other hand. These findings raise a question: does the critical probability, $p_c$, that marks the measurement-induced phase transition exhibit a similar relationship with the above quantities? For instance, the maximum entanglement is sustained for the case when $\mathcal{U}=\text{iSWAP}$, as shown in the Fig.~\ref{fig5}. Does this case remain the most robust to measurements in having the largest value of $p_c$?




From Eq. \eqref{eq1}  it is clear that we need three parameters to describe a two-qubit gate or its LU equivalent. However, the $e_p -g_t$ plane represents a two parameter family and there exists gates that have the same $e_p - g_t$ value but are not LU equivalent in general.
Nonetheless, as we have seen with the behavior of entanglement entropy, the study of MIPT in the $e_p -g_t$ plane gives us valuable insights and sheds light on the operational role of these quantities in critical behavior.
Only in the case of gates lying along the boundary, i.e. gates from SWAP to iSWAP and $\mathcal{I}$ to SWAP, one can uniquely specify them by their $e_p -g_t$ values. 
 In Table~\ref{tab:epgtpcnu} we list the $p_c$ and $\nu$ for the gates along the four boundaries in Fig.~\ref{fig3} using 6 different values for the system size $L \in [6,8,10,12,14,16]$. The critical probability increases as $e_p$ increases along the T-dual line.  Along this line, numerically we observe $\nu$ values clustered near 1.5 and 2.7 with an error of $0.2$, indicating the existence of at least two universality classes.
\begin{table}[b]
\caption{\label{tab:epgtpcnu}%
Critical probability and exponent for two-qubit unitaries along the boundaries of the $e_p - g_t$ plane. Errors in the $p_c$ and $\nu$ values are around $10\%$ and $13\%$ respectively.}
\begin{ruledtabular}
\begin{tabular}{cdddd}
  & 
\multicolumn{1}{c}{$e_p$}&
\multicolumn{1}{c}{$g_t$}&
\multicolumn{1}{c}{$p_c$}&
\multicolumn{1}{c}{$\nu$}\\
\hline

\rowcolor{mixColor4}
$\mathcal{I}$ & 0.0 & 0.0 & 0.049 & 0.99\\
\rowcolor{blue!20} 
 & 0.074 & 0.037 & 0.09& 1.5\\
\rowcolor{blue!20}
& 0.222 & 0.111 & 0.15 & 1.7\\
\rowcolor{blue!20}
\rowcolor{blue!20} 
T-dual & 0.370 & 0.185 & 0.32 & 2.7\\
\rowcolor{blue!20}
& 0.444 & 0.222 & 0.38 & 2.6\\
\rowcolor{blue!20}
& 0.518 & 0.259 & 0.36 & 2.0\\
\rowcolor{blue!20}
& 0.592 & 0.296 & 0.42 & 2.0\\
\rowcolor{mixColor} 
CNOT & 0.667 & 0.333 & 0.367 & 1.50\\
\rowcolor{red!20}
 & 0.667 & 0.370 & 0.36 & 1.4\\
\rowcolor{red!20}
 & 0.667 & 0.407 & 0.40 & 1.4\\
\rowcolor{red!20}
 & 0.667 & 0.444 & 0.41 & 1.4\\
\rowcolor{red!20}
Maximum $e_p$ & 0.667 & 0.481 & 0.37 & 1.2\\
\rowcolor{red!20}
 & 0.667 & 0.518 & 0.40 & 1.2\\
\rowcolor{red!20}
 & 0.667 & 0.555 & 0.39 & 1.2\\
\rowcolor{red!20}
 & 0.667 & 0.592 & 0.39 & 1.1\\
\rowcolor{red!20}
 & 0.667 & 0.629 & 0.37 & 1.0\\
\rowcolor{mixColor1}
iSWAP & 0.667 & 0.667 & 0.388 & 1.06\\
\rowcolor{yellow!20} 
  & 0.592 & 0.704 & 0.4 & 1.1\\
\rowcolor{yellow!20}
 & 0.518 & 0.741 & 0.39 & 1.1\\
\rowcolor{yellow!20}
 & 0.444 & 0.778 & 0.40 & 1.2\\
\rowcolor{yellow!20}
Dual-unitary& 0.370 & 0.815 & 0.36 & 1.1\\
\rowcolor{yellow!20}
 & 0.222 & 0.889 & 0.40 & 1.6\\
\rowcolor{yellow!20}
 & 0.074 & 0.963 & 0.34 & 2.0\\
 \rowcolor{mixColor3}
 SWAP & 0.0 & 1.0 & 0.05 & 1.2 \\
\rowcolor{green!20}
  & 0.173 & 0.904 & 0.36 & 1.2 \\
\rowcolor{green!20}
 $\text{SWAP}^\alpha$& 0.5 & 0.5 & 0.37 & 1.0 \\
\rowcolor{green!20}
 & 0.173 & 0.095 & 0.15 & 2.4 \\
\end{tabular}
\end{ruledtabular}
\end{table}

We find that the gates with maximum entangling power, along the CNOT-iSWAP line,  
have near maximal $p_c$ value of $0.37$ as shown in the Table~\ref{tab:epgtpcnu}
 For the CNOT and iSWAP gates, we considered a system size up to 20 qubits, which increases the accuracy in $p_c$. The critical probability $p_c$ is 0.367 for CNOT and $0.388  \pm 0.002$ for iSWAP, illustrating the role of $g_t$ in MIPT. This observation is further strengthened when we consider two points with the same $e_p = 0.173$ on the parabolic boundary corresponding to $\text{SWAP}^\alpha$, one with $g_t = 0.904$ and the other with $g_t = 0.095$, we observe a clear difference in $p_c$ as shown in Table~\ref{tab:epgtpcnu}. Therefore, we can conclude that for small values of $e_p$, $g_t$ is crucial in determining $p_c$. For the $\mathcal{I}$ and SWAP gates, a non-zero $p_c$ shown in Table~\ref{tab:epgtpcnu} is due to finite system size error. For large $N$, the binomial distribution for the number of unmeasured qubits in Eq.~(\ref{partaAprob}) may be approximated by the normal distribution for a fixed value of $p$
 and $t=4N$. As $(1-p)^t \approx e^{-4Np}$, the average number of unmeasured qubits in either part $A$ or $B$ goes as 
 $N e^{-4Np}$. Thus except when $p=0$, the probability of having unmeasured qubits vanishes. 
Consequently, when $p=0$, for the case of the $\mathcal{I}$ and SWAP gates, the entanglement will scale as the volume law (as, we are starting from a Haar random state and $S(N_A:N_B) \approx N$), but this entanglement will vanish for any $p>0$ as $\approx N e^{-4Np}$.

\begin{figure}[!htbp]
\centering
\includegraphics[width=1\linewidth]{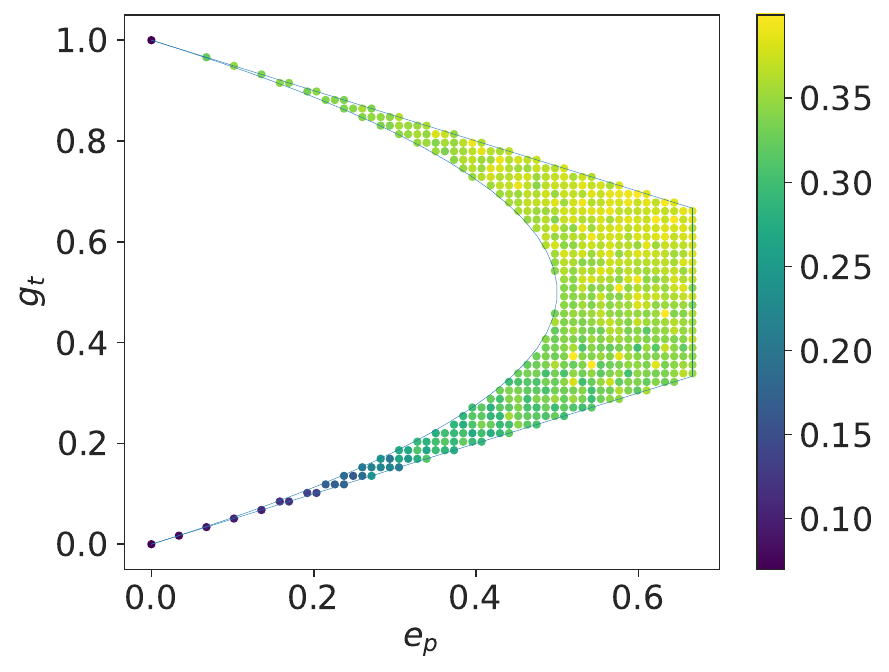}
\caption{Classification of unitaries in the $e_p-g_t$ plane with respect to critical probability $p_c$ at which the phase transition occurs. Color bars show the $p_c$ values. In the vicinity of the $\mathcal{I}$-CNOT region, $g_t$ is bellow 0.5, the critical probability increases with $e_p$. The gates near the iSWAP gate exhibit a significantly higher critical probability, approximately 0.36. The figure is obtained by uniformly sampling 614 unitaries from the interior of the permissible region in the $e_p-g_t$ plane and calculating the critical value $p_c$ at which the phase transition occurs.}
\label{fig10}
\end{figure}

 We now explore the critical parameters in the interior region of Fig.~\ref{fig3}. To dive into this inquiry, we select unitaries uniformly from the $e_p-g_t$ plane and repeat the calculations similar to the one for the CNOT gate in the previous section. 
 We have sampled 614 unitaries $\mathcal{U}$ uniformly spread across the territory outlined in Fig.~\ref{fig3}.
 With equations Eq.~\eqref{eq:epgt_formulas}, we calculate $\{c_1, c_2, c_3\}$ for each unitary with a particular $e_p$ and $g_t$ value. Once we get $\{c_1, c_2, c_3\}$, we carry out circuit evolution, measurements on qubits, calculating bipartite entanglement, etc. We calculate the entropy curve for each point in the $e_p$-$g_t$ plane for different system sizes (as in Fig.~\ref{fig6}). 
 To calculate the critical $p_c$ and the critical exponent $\nu$ we use finite size scaling outlined in the previous section.

 The critical probability value $p_c$ for various values of $e_p$ and $g_t$ is illustrated in Fig.~\ref{fig10}. A notable observation emerges: the interior region approximately interpolates the $p_c$ values observed along the boundaries. We also observe gates positioned near the lower values of $e_p - g_t$ exhibit significantly smaller values of $p_c \approx 0.1$, denoted by dark blue dots in Fig.~\ref{fig10}.  These gates are positioned in close proximity to the identity gate $\mathcal{I}$ in the $e_p - g_t$ plane and showcase minimal entangling power.  This observation strengthens our claim that gates possessing very low entangling power can lead to a phase transition with a relatively low probability of measurement in a hybrid circuit.  Another notable region corresponds to the maximum $p_c \approx 0.39$, (region colored yellow in Fig.~\ref{fig10}). For gates near iSWAP, the entanglement in hybrid circuits shows enhanced resilience to measurements. For instance, even if approximately $30\%$ of the total qubits are measured after each evolution, entanglement tends to exhibit characteristics indicative of long-range entanglement adhering to a volume law. This region opens up intriguing avenues for future exploration on the interplay of global bipartite entanglement and sustaining it through
only two-qubit entangling gates despite appreciable measurement probability. 

At the other end of the dual line from the iSWAP is the SWAP gate, and we see from Table~\ref{tab:epgtpcnu} that $p_c$ vanishes
for this, while for gates in the vicinity there is a substantial value of $p_c$. It has been shown that arbitrary close to the SWAP gate, the circuits are mixing and have random matrix properties in the thermodynamic limit \cite{Bertini_2021CMP}. Thus one may expect that except for the SWAP gate, the circuits in the vicinity are generic.
This is in contrast to the Identity gate where the $p_c$ also vanishes. Both the identity and the SWAP gate generate no entanglement and as soon as all qubits are measured, the circuit will have zero entanglement. 
Hence even a vanishingly small $p_c$ is enough to destroy the entanglement in the circuit.


\begin{figure}[!htbp]
\centering
\includegraphics[width=1\linewidth]{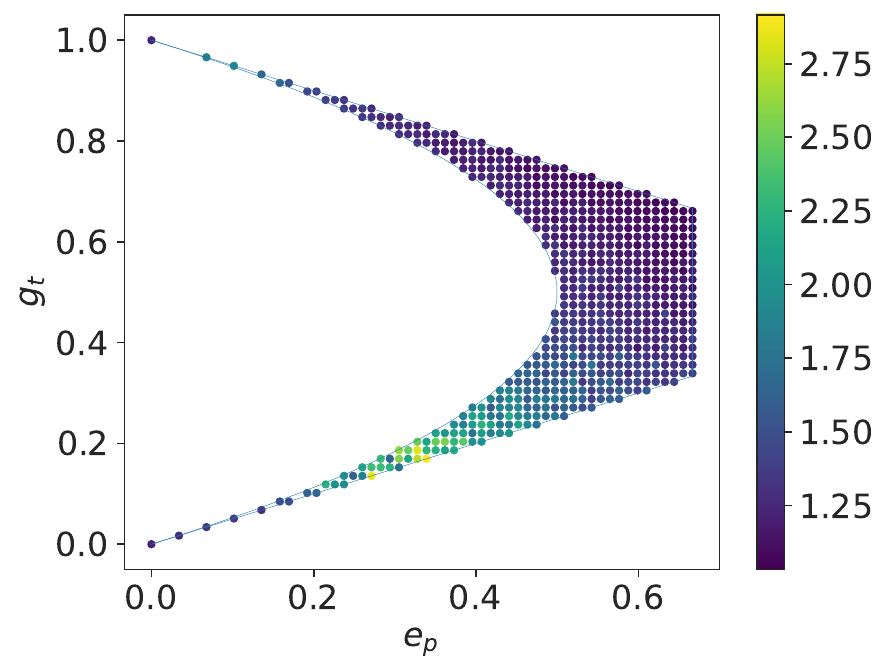}
\caption{Classification of unitaries in the $e_p-g_t$ plane with respect to critical exponent, $\nu$, at which the phase transition occurs. Color bars show the $\nu$ values. The figure is obtained by uniformly sampling 614 unitaries from the interior of the permissible region in the $e_p-g_t$ plane and calculating the critical exponent $\nu$ corresponding to the phase transition.}
\label{fig11}
\end{figure}





\section{Discussion}
The ``quantum" element of  quantum information processing can be relevant at any of the three stages of a quantum process - initial preparation of highly entangled non-classical states, unitary evolution given by the Schr\"{o}dinger's equation and the collapse of the wave function through quantum measurements. All three stages are fundamental ingredients of any quantum circuit implementing a quantum information processing protocol. Examples include quantum teleportation and dense coding, the question of efficient classical simulation of such circuits, thermalization using t-designs and other applications of random quantum circuits implementing quantum algorithms and measurement-based quantum computation\cite{briegel2009measurement, nielsen2001quantum, PhysRevA.80.012304}. From a foundational point of view, the emergence of classical physics from underlying quantum mechanics, the system apparatus interaction giving rise to pointer states and decoherence are an interplay of evolution and measurement\cite{RevModPhys.75.715}. 
 A hybrid circuit consisting of entangling unitaries and measurements also serves as a toy model for interacting multi-partite complex quantum systems usually having a large Hilbert space and undergoing local projective measurements.

In this work, we investigate the behavior of entanglement entropy and measurement induced phase transition in a quantum circuit with a unitary of definite entangling power and gate typicality interspersed with random local measurements with a certain probability. We took a circuit model where locals are Haar random, but the non-local element is the same two-qubit gate applied throughout the circuit. 
We focus on the amount of bipartite entanglement present between the two halves of a hybrid circuit that has both entangling two-qubit operations followed by local projective measurements after a sufficient amount of time. We then calculate the entropy as a function of the probability of measurement, $p$, by averaging over many circuit realizations with the same non-local component $\mathcal{U}$. We find that, for a fixed value of gate typicality, the entanglement entropy is positively correlated with the entangling power. For the case when the non-local component $\mathcal{U}$ is the identity, we obtain analytical estimate for the entanglement entropy and show an excellent agreement with numerical simulations. 

Furthermore, we show that for a definite entangling power, the entanglement entropy in the circuit is positively correlated with gate typicality. This gives another way of interpreting gate typicality by associating it with a concrete and well-defined physical quantity like the bipartite entanglement in a circuit. 

We show the existence of MIPT for the hybrid circuit model for the entire spectrum of two-qubit unitary gates. We find that entangling power as well as gate typicality play a crucial role in the behavior of entanglement entropy under measurement. 
We also find that the entangling power and gate typicality can be used to classify two-qubit unitaries into different universality classes of phase transitions that can occur in a hybrid circuit. In particular, we discover there are atleast two distinct university classes for the MIPT.

Future directions include the study of hybrid quantum circuits with entangling unitaries comprising of quantum chaotic maps with a variable chaoticity parameter giving rise to regular, mixed or globally chaotic dynamics. One can then ask what is the entanglement behavior in such circuits and also the nature of phase transitions that can be induced. An approach based on renormalization group analysis can give us a better understanding of the nature of phase transitions. Two-qubit gates, apart from the entangling power and gate typicality, have a third independent invariant. Its operational interpretation and physical significance is unclear, but it will be of interest to study its influence on MIPT.
Lastly, while we have given a phase diagram for measurement-induced phase transitions for two-qubit, gates in higher dimensions are completely uncharted territory. 
\section{Acknowledgements}
	
	 The authors would like to thank HPCE, IIT Madras, for providing the computational facility for numerical simulations. 
  This work was supported in part by grant  DST/ICPS/QusT/Theme-3/2019/Q69 and New faculty Seed Grant from IIT Madras. The authors were supported, in part, by a grant from Mphasis to the Centre for Quantum Information, Communication, and Computing (CQuICC) at IIT Madras.



\bibliography{main}

\end{document}